\documentclass[conference]{IEEEtran}
\IEEEoverridecommandlockouts
\usepackage{cite}
\usepackage{amsmath,amssymb,amsfonts}
\usepackage{algorithmic}
\usepackage{graphicx}
\usepackage{textcomp}
\usepackage{xcolor}
\usepackage{listings}
\usepackage{makecell}
\usepackage{float}
\usepackage{tabularx}

\def\BibTeX{{\rm B\kern-.05em{\sc i\kern-.025em b}\kern-.08em
    T\kern-.1667em\lower.7ex\hbox{E}\kern-.125emX}}
\begin{document}

\title{Not All Nudges Land: Behavioral Controllability and Elaboration Quality in AI-Supported Journaling\thanks{\textcolor{red}{Accepted and presented at the HARMONY Workshop, IEEE/ACM Conference on Connected Health: Applications, Systems and Engineering Technologies (CHASE 2026)}}}
\author{
\IEEEauthorblockN{ Nadia Mehjabin}
\IEEEauthorblockA{\textit{Department of Computer Science} \\
\textit{University of Virginia}\\
Charlottesville, USA \\
jqc7gj@virginia.edu}
\and
\IEEEauthorblockN{ Henry Kautz}
\IEEEauthorblockA{\textit{Department of Computer Science} \\
\textit{University of Virginia}\\
Charlottesville, USA \\
rmw7my@virginia.edu}
\and
\IEEEauthorblockN{ Subigya Nepal}
\IEEEauthorblockA{\textit{Department of Computer Science} \\
\textit{University of Virginia}\\
Charlottesville, USA \\
sknepal@virginia.edu}
}
\maketitle

\begin{abstract}
AI journaling tools can tailor prompts to a person's own sensed behavior, but it is unclear which behaviors respond to them. We analyzed 369 journal entries from an eight-week passive sensing study. An LLM labeled each entry as expressing an intention to change a behavior or not, and we measured follow-through against 26 sensor features with a 3-day before/after comparison. Responsiveness depended most on whether a behavior involves other people. Behaviors that depend on others improved in only 15 to 22\% of cases, while behaviors a person can act on alone improved more often, up to 50 to 63\%, though unevenly. How users wrote mattered less. No single text feature separated improved from unimproved entries; writing carried signal only within specific behaviors, most clearly for text messaging and for longer, more personal intention entries. The sample is small, so we treat these as exploratory patterns that point to where AI journaling nudges are most likely to work.
\end{abstract}


\section{Introduction}

Behavioral sensing systems that integrate passive data collection with AI-generated prompts or nudges represent an emerging approach to support health behavior change. Passive sensing has been used to personalize health behavior interventions, and large language models (LLMs) have been applied to support journaling and self-reflection. Recent work has begun combining these two approaches, using continuous sensing data to generate LLM-based journaling prompts grounded in users' own behavioral context. We conducted an 8-week study of the MindScape system with undergraduate students and demonstrated significant gains in well-being~\cite{nepal2024mindscape}. Yet, our system distributes prompts equally across all behavioral domains based on sensing data availability rather than behavioral responsiveness.

In an initial analysis of the MindScape \cite{nepal2024mindscape} dataset, we manually examined journal entries for statements of intent and compared them to subsequent behavioral sensing data. Several cases showed clear alignment: a participant who wrote about wanting to walk more showed increased walking episodes in the following days; another who mentioned reaching out to friends showed higher SMS counts in the same period. These patterns were observable but not measurable; we had no systematic way to identify intention entries at scale, and no method to statistically connect journal content to sensing outcomes across the full dataset. This led us to build an LLM-based classification pipeline for this task, paired with a before/after sensing comparison to test follow-through at scale.

Applying this pipeline to 369 journal interactions across 26 sensing features, we find that responsiveness depends mostly on whether a behavior involves other people: behaviors that depend on social context or other people improve in only 15 to 22\% of cases, while behaviors a person can act on alone improve more often (up to 50 to 63\%) but inconsistently. For text messaging, the most immediately actionable behavior, intention language was linked to improvement in a small set of entries. Within intention entries, richer elaboration was also linked to follow-through. We make two contributions. First, a responsiveness map across the 26 sensing features, in which dependence on other people -- more than individual controllability -- separates responsive from unresponsive behaviors. Second, descriptive evidence that, in this sample, two things track short-term follow-through: intention language for individually controllable behaviors such as SMS, and richer elaboration within intention entries. Together, these help designers decide where to target nudges and what in a user’s response signals likely follow-through.

\section{Related Work}

Journaling supports self-awareness \cite{alt2020reflective, williams2009reflective}, emotional processing \cite{smyth2018positiveaffectjournaling}, and cognitive organization \cite{sohal2022efficacy},
with evidence showing benefits for mood, stress reduction, and
mental well-being \cite{keech2021journaling, miller2014interactive, sohal2022efficacy}. A key recent advance has been the shift from generic journaling prompts toward context-aware prompting grounded in users' own behavioral data. MindScape~\cite{nepal2024mindscape} 
exemplifies this direction, combining passive sensing with large language models to deliver personalized journaling prompts, demonstrating meaningful well-being gains over 8 weeks. Similar systems including MindfulDiary~\cite{kim2023mindfuldiary} and Reflection Companion~\cite{kocielnik2018reflection} have shown that AI-mediated journaling can support 
consistent reflective habits. Related work on daily planning prompts has likewise shown that prompting users to form concrete plans can shape routine behaviors ~\cite{cuadra2021planning}. However, MindScape distributes prompts across behavioral domains based on sensing availability, without accounting for whether those domains are equally responsive to nudges. A large-scale meta-analysis of over 200 studies found that while nudges produce reliable behavior change overall, effectiveness varies substantially by domain~\cite{mertens2022effectiveness}.
Personal informatics research has likewise found that self-monitoring and reflection do not affect all health behaviors equally. Frameworks for mobile behavior change stress that responsiveness depends on the behavior, the moment, and the person together~\cite{okeke2018towards}. Physical activity responds reliably to step-count feedback and goal-setting nudges, while sleep and social behaviors are harder to shift through brief app-based interventions~\cite{bakker2018engagement}. Bhattacharjee 
et al.~\cite{bhattacharjee2023integrating} further showed that behaviors embedded in social 
coordination are less amenable to individual-level 
prompting than behaviors under personal control. To the best of our knowledge, no prior study has empirically characterized which domains respond to journaling nudges and which do not.

Even when a nudge reaches the right domain, whether it produces change depends on how the user responds. Gollwitzer's work on implementation intentions showed that specificity is the critical mediator: concrete plans predict follow-through more reliably than vague expressions of intent~\cite{gollwitzer1999implementation}. Prior journaling work has analyzed content for sentiment and linguistic patterns \cite{nepal2024mindscape, 
kim2023mindfuldiary}, but has not connected the quality of users' responses to subsequent behavioral 
outcomes. Englhardt et al.~\cite{englhardt2024clinicalinsights} showed that LLMs can reason over mobile behavioral data to generate clinical insights, but to our knowledge LLM-based intention classification has not been paired with before/after sensing validation. The pipeline developed in this paper addresses both gaps.

\section{Methods}

This study is a secondary analysis of the MindScape longitudinal study \cite{nepal2024mindscape}. Full details of the study design, participant recruitment, app architecture and prompt generation pipeline are described in our prior papers ~\cite{nepal2024contextual, nepal2024mindscape}. We summarize the aspects most relevant to our analysis here. 

\subsection{Dataset}\label{AA}

The MindScape study enrolled 20 undergraduate students at Dartmouth College over 8 weeks. During Weeks 1 through 6, participants received contextual AI-generated prompts derived from their passive sensing data. During Weeks 7 through 8, participants received generic prompts not grounded in behavioral data. Our analysis focuses on the first 6 weeks of contextual journaling, yielding 20 participants and 369 journal entries.

\begin{table}[ht]
\centering
\caption{Passive Sensing Features per Behavioral Domain}
\label{tab:features}
\begin{tabular}{lp{5.4cm}c} 
\hline
\textbf{Domain} & \textbf{Features} & \textbf{N} \\
\hline
Digital Habits & Communication, Entertainment, and Social app use; Screen unlock duration and count & 5 \\
\noalign{\smallskip} 
Social Interaction & Call duration/count (In/Out), Conversation duration/count (Food venues, Home), Time at specific locations (Greek spaces, Dorms, Study), SMS count (In/Out) & 16 \\
\noalign{\smallskip}
Sleep & Sleep duration & 1 \\
\noalign{\smallskip}
Physical Fitness & Gym/workout duration; Cycling, Walking, and Running episode durations & 4 \\
\hline
\textbf{Total} & & \textbf{26} \\
\hline
\end{tabular}
\end{table}

\subsection{Behavioral Signal and Improvement Operationalization}
For each of the four behavioral domains, physical fitness, sleep, digital habits, and social interaction, we identified sensing features representing the full 24-hour daily aggregate as the primary behavioral signal. The features included per domain are shown in Table ~\ref{tab:features}.

For each journal entry, we identified the behavioral features most relevant to the journal content using the following procedure. First, we searched the journal response text for explicit or implicit references to a sensing feature. For example, a mention of communication or texting was mapped to \textit{app\_Communication} and \textit{sms\_in\_num}. If no feature was identifiable from the response, we examined the prompt text to determine the behavioral category and associated features. When multiple features were referenced, we selected the feature with the maximum absolute signal change as the primary improvement indicator for that entry.

To measure whether a journal entry corresponded to behavioral change, we computed the mean value of each sensing feature over the 3 days before and 3 days after each entry. We evaluated 1, 3 and 7-day windows and selected the 3-day window as the primary analysis window. The overall improvement rates across the three windows are 43.8\% (1-day), 45.8\% (3-day), and 47.3\% (7-day), a difference of only 3.5 percentage points across the full range. Domain-level rates are similarly stable, meaning the choice of window does not meaningfully change the pattern of results.

We then assigned a binary improvement flag: 1 if the 3-day post-entry mean exceeded the 3-day pre-entry mean in the direction of improvement for that feature, 0 otherwise. The improvement direction was feature-specific and domain-informed: for example, we coded a decrease in phone usage duration as improvement for digital habits, while we coded an increase in walking duration as improvement for physical fitness. This yielded 26 features with valid improvement flags across 369 entries, which form the basis of all subsequent analyses.

\subsection{LLM Classification Pipeline}

\begin{figure}[t]
  \centering
  \includegraphics[width=0.85\columnwidth]{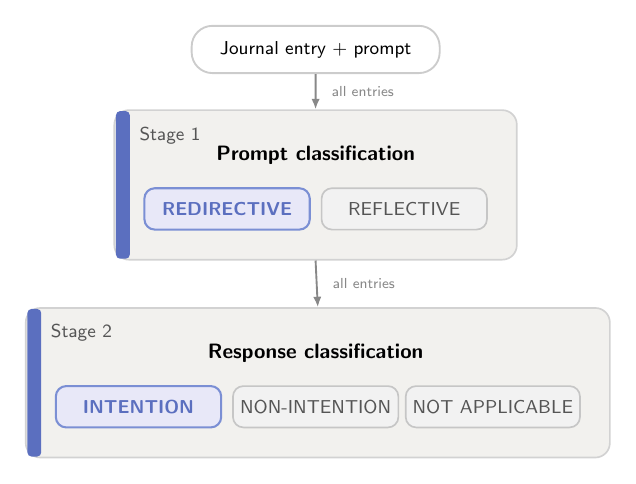}
  \caption{Two-stage LLM classification pipeline: Stage 1 classifies each system prompt as REDIRECTIVE (behavior-targeted) 
  or REFLECTIVE (open-ended); Stage 2 classifies each journal response as INTENTION (a stated plan or commitment), NON-INTENTION (reflection without commitment), or NOT APPLICABLE.}
  \label{fig:pipeline}
\end{figure}

Our initial exploration of the dataset used a time-series language model to detect intention language in journal entries at scale. That approach performed poorly, failing to reliably distinguish entries expressing behavioral intentions from general reflections. This motivated the development of a two-stage LLM classification pipeline (Fig. ~\ref{fig:pipeline})  using \texttt{llama-3.3-70b-versatile}.

\textbf{Stage 1 - Prompt classification:} Each system prompt was classified as REDIRECTIVE (explicitly targeting a behavioral domain and inviting the user to set a goal or plan) or REFLECTIVE (open-ended, inviting general reflection without a specific behavioral target).
\textbf{Stage 2 - Response classification:} Each journal entry paired with its prompt was classified as INTENTION (the user expressed a plan, goal, or commitment to change a behavior), NON-INTENTION (the user reflected on past behavior without committing to change), or NOT APPLICABLE (the entry did not engage with the prompt's behavioral focus).

To validate the pipeline, we independently labeled a random sample of 30 entries, obtaining an aggregate agreement rate of approximately 70\% across prompt and response classifications. This study is exploratory, and telling reflective from redirective prompts (and intention from non-intention responses) is partly subjective. We treat this agreement as adequate to proceed, note it as a limitation and encourage future work to build purpose-built annotation schemes.

\subsection{Linguistic Feature Extraction}
To examine whether the linguistic properties of journal entries predict behavioral follow-through, we extracted features at three levels.
\textbf{Surface features:} We computed word count and type-token ratio (TTR) for each entry. TTR measures lexical diversity: a lower TTR in intention entries indicates more focused, repetitive vocabulary centered on a specific behavior, which we hypothesized would be associated with follow-through.
\textbf{Syntactic features:} Using dependency parsing and morphological analysis, we extracted first-person singular pronoun count (I, me, my, myself) as a proxy for personal agency and commitment, and future orientation as a measure of forward-looking language.
\textbf{LLM semantic scores:} Each entry was scored by the LLM on three dimensions: behavioral concreteness (1 to 4, from no specific behavior mentioned to specific behavior with clear timing), planning depth (1 to 3, from awareness only to concrete plan), and emotional engagement (1 to 3, from detached to strong personal investment). We used these scores to check whether the INTENTION versus NON-INTENTION classification reflected qualitative differences in entry content, and whether semantic quality within INTENTION entries further predicted improvement. The full rubric and prompt are in the Appendix (Table ~\ref{tab:system_v2}). 

\section{Results}
\subsection{Pipeline Validation}
As shown in 
Table~\ref{tab:prompt_response}, REDIRECTIVE prompts elicited 
INTENTION responses in 49.5\% of cases, compared to only 9.3\% 
for REFLECTIVE prompts (Fisher's exact p~$<$~0.01). This pattern is expected, since redirective prompts explicitly invite users to set goals or make plans while reflective prompts invite open-ended reflection. Because the classifier picks up this difference, it appears to capture a real linguistic distinction. This makes the INTENTION versus NON-INTENTION comparisons that follow meaningful.

\begin{table}[h]
\centering
\caption{Response type distribution by prompt type. 
REDIRECTIVE prompts elicit intention responses at a 
significantly higher rate than REFLECTIVE prompts 
(Fisher's exact p~$<$~0.0001).}
\label{tab:prompt_response}
\begin{tabular}{lcccr}
\hline
\textbf{Prompt type} & \textbf{INTENTION} & 
\textbf{NON-INTENTION}  & 
\textbf{Intention rate} \\
\hline
REDIRECTIVE  & 50 & 48   & 49.5\% \\
REFLECTIVE   & 25 & 223  & 9.3\%  \\
\hline
\end{tabular}
\end{table}

\subsection{Finding 1: Social Dependence, More Than Individual Controllability, Predicts Responsiveness}

Improvement rates in 26 sensing features (see Table \ref{tab:feature_ranking}) reveal a pattern that is clearest at the extremes and tied most reliably to whether a behavior depends on other people.

Behaviors primarily driven by individual decisions tended to improve more often: incoming SMS (62.9\%), walking episodes (58.0\%), and time 
spent at home (53.6\%) improved in over half of cases. Behaviors involving some scheduling friction or environmental dependency  
improved in 40 to 50\% of cases: phone usage duration (48.2\%), running (43.5\%), and communication 
app usage (42.5\%). Behaviors most dependent on social coordination or physical co-presence with others rarely 
improved: phone call features (25 to 31\%), food venue 
conversations (22\%), and Greek space visits (15\%). 

The low end is the most consistent part of this pattern: behaviors that require other people's coordination, such as calls, food-venue conversations and Greek-space visits, were uniformly among the least responsive. The high end is noisier, and individual controllability did not by itself predict improvement. Among behaviors a person performs alone, walking improved in 58.0\% of cases but gym/workout time in 20.0\%, entertainment-app use in 18.8\%, and cycling in 12.5\% — near or below the socially dependent behaviors. Texting shows the same split: incoming SMS improved in 62.9\% of cases but outgoing SMS in only 38.5\%. We therefore read this result less as a clean controllability gradient than as a single robust asymmetry: behaviors that depend on other people are consistently unresponsive to journaling nudges, whereas behaviors under individual control vary widely and are not explained by controllability alone. Because we grouped behaviors by judgment rather than from a separate measure of controllability, and several features rest on small samples, we treat the ordering as descriptive.

\subsection{Finding 2: Intention Language Is Associated With Outcomes for Highly Responsive Behavior in a Small Exploratory Sample}

As noted earlier, an overall test across all entries found no link between text features and improvement. So we treat this subsection and the next as exploratory patterns within specific behaviors, meant to generate hypotheses. Among individual sensing features, SMS communication showed the clearest relationship between response type and behavioral outcome. For incoming SMS (or text messages), entries classified as INTENTION were followed by improvement in 100\% of cases (8 of 8), compared to 52.2\% for NON-INTENTION entries (Fisher's exact p~=~0.028). These 8 entries came from 7 different participants, so the result is not just one person repeated. Still, it rests on very few entries and should be read with caution. Outgoing SMS showed a directionally consistent pattern, with INTENTION entries exhibiting larger behavioral change than NON-INTENTION entries (Mann-Whitney U, p =~ 0.046).

SMS is a plausible case for this association because it is immediately actionable: deciding to reach out to someone and sending a message can happen in the same moment as writing a journal entry. This sits at the most responsive end of the behaviors observed in the prior finding. We note, however, outgoing SMS is arguably more directly controllable than incoming SMS, as the sender can initiate contact unilaterally. The stronger result for incoming SMS may mean that journaling about social intentions makes people more open to others’ messages — responding and being available — rather than starting contact themselves. Both directions showed consistent patterns (incoming: Fisher's exact p = 0.028; outgoing: Mann-Whitney U p = 0.046), though the small sample warrants caution in any mechanistic interpretation.

\subsection{Finding 3: Elaboration Quality Within Intentions 
Is Associated With Follow-Through}

Intention language alone is not sufficient to predict follow-through. Within INTENTION entries specifically, the quality of elaboration is further associated with whether behavioral improvement occurs. Restricting analysis to INTENTION entries only (improved: n~=~27, not improved: n~=~44), three linguistic features differentiated the two groups.

\begin{figure*}[t]
    \centering
    \begin{minipage}{0.32\textwidth}
        \centering
        \includegraphics[width=\textwidth]{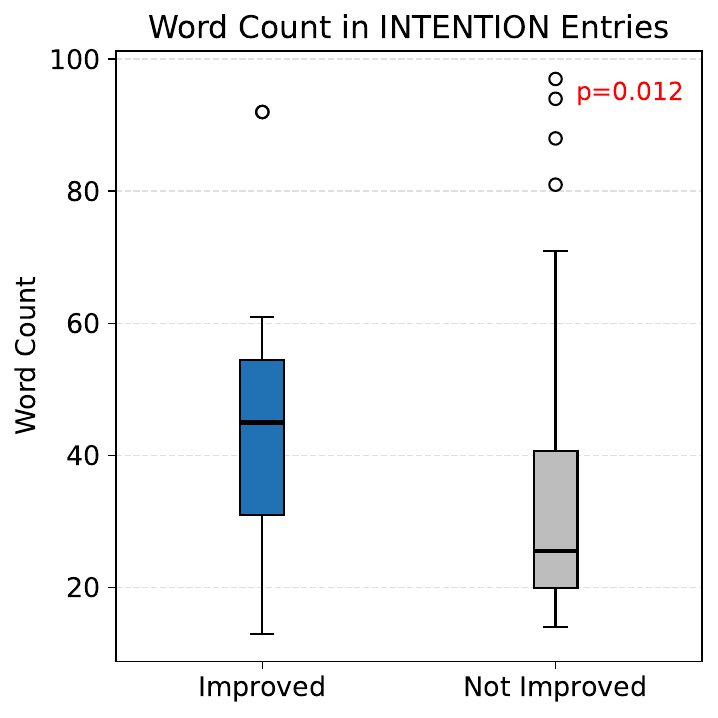}
    \end{minipage}
    \hfill
    \begin{minipage}{0.32\textwidth}
        \centering
        \includegraphics[width=\textwidth]{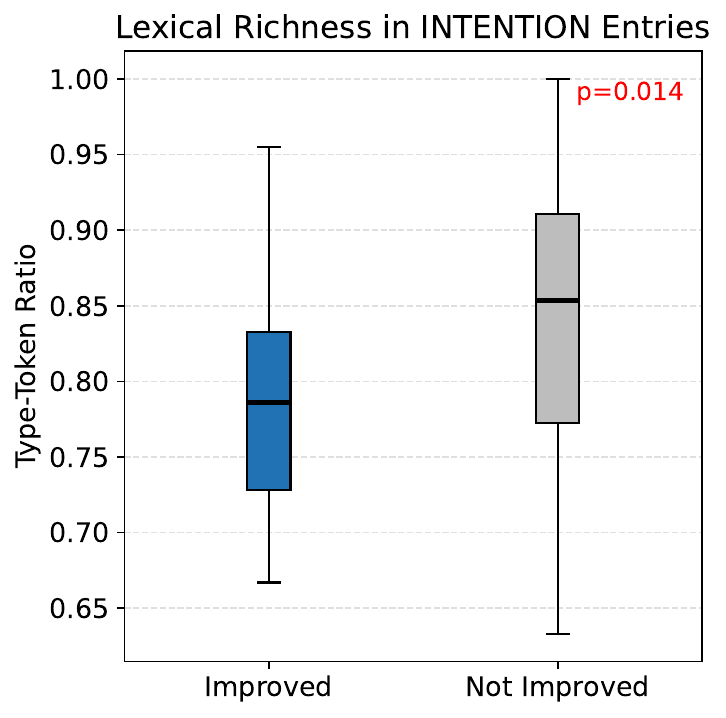}
    \end{minipage}
    \hfill
    \begin{minipage}{0.32\textwidth}
        \centering
        \includegraphics[width=\textwidth]{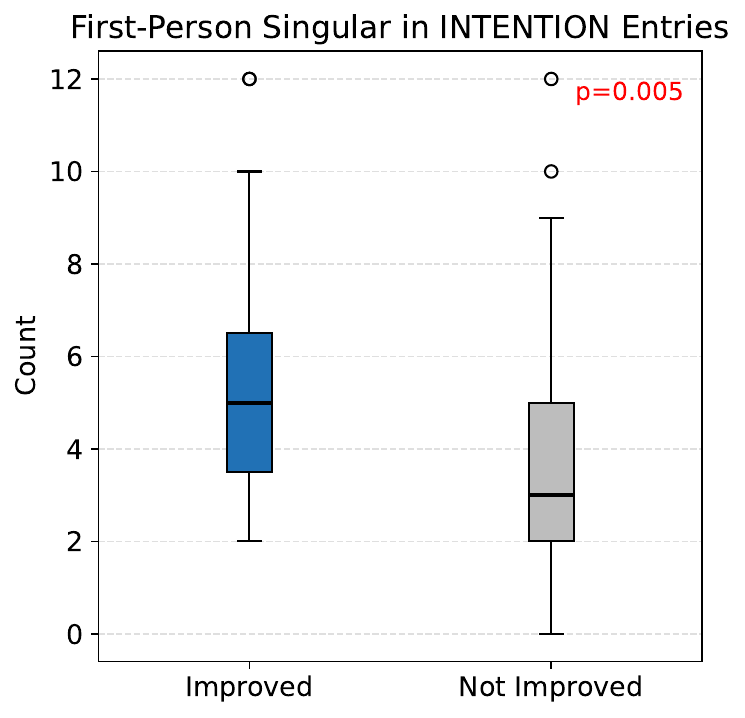}
    \end{minipage}
    \caption{Word count and lexical richness (TTR) within 
  INTENTION entries, split by behavioral outcome (improved 
  n~=~27, not improved n~=~44). Improved entries are 
  significantly longer (median 45 vs. 25.5, 
  p~=~0.012) and show lower TTR indicating more focused, 
  repetitive vocabulary centered on a specific behavior 
  (median 0.79 vs. 0.85, p~=~0.014).}
    \label{fig:Data}
\end{figure*}

Improved intention entries were longer (Fig. ~\ref{fig:Data}, median word count 45 vs. 25.5, Mann-Whitney p~=~0.012, r~=~0.21). They also had lower type-token ratios indicating more focused vocabulary centered on a specific behavior (TTR 
median 0.79 vs. 0.85, p~=~0.014, r~=~$-$0.26). But since improved entries were also longer, this lower diversity partly just reflects length, because TTR falls as texts get longer. They also used more first-person singular language (Fig. ~\ref{fig:Data}, median count 5 vs. 3, p$<$0.01). Short, vague entries were not linked to behavioral change; longer, more personal ones were. 

LLM semantic scoring showed that INTENTION entries differ from NON-INTENTION entries in behavioral concreteness, planning depth, and emotional engagement  (all p~$<$~0.05), consistent with the classifier capturing qualitative differences in content. Within INTENTION entries, though, these scores did not separate improved from not-improved entries. This suggests the signal lies in how much someone elaborates and in their personal voice, not in the topic or category.

\section{Discussion}
\subsection{Relationship to the Original MindScape Findings}

In the main MindScape paper, we asked whether AI journaling improves well-being. In this secondary analysis, we ask which interactions within it drive change. Across all 369 entries, no single text feature predicted behavioral improvement (all p~$>$~0.10), and intention and non-intention entries did not differ between the improved and unimproved groups ($\chi^2$~p~=~0.15). 

This interaction-level null fits the study-level well-being gains that the main paper reported~\cite{nepal2024mindscape}. Participants self-reported those gains, which built up over eight weeks of steady journaling, while we measure short-term behavioral change in passive sensing around each entry. A behavior does not have to move in the three days after an entry for journaling to help over two months. Our analysis adds resolution. The interactions that most often preceded near-term change involved behaviors a user can carry out alone and intentions the user elaborated, while behaviors that depend on other people rarely changed. The study-level result leaves this pattern implicit.

\subsection{Theoretical Grounding}
The most robust part of our result, that socially dependent behaviors resist change, lines up with two established bodies of theory. Self-determination theory~\cite{deci2000and} distinguishes between autonomously regulated behaviors: those driven by internal motivation and executable through individual decision, and those 
requiring external coordination. Food-venue conversations and Greek-space events require other people’s presence and cooperation, and these externally coordinated behaviors were consistently the least responsive, as the theory would predict. The autonomous end was less clean: walking responded strongly, but other self-directed behaviors such as cycling did not, so the theory explains the unresponsive floor better than it predicts the responsive ceiling.

The elaboration finding echoes Gollwitzer's work on implementation intentions~\cite{gollwitzer1999implementation}: specific, concrete plans are more reliably followed through than vague expressions of intent. Our word count and first-person singular results mirror that pattern in  naturalistic journaling data, extending a laboratory finding into a real-world mobile health context.

\subsection{Design and Research Implications}
These findings suggest several directions for both practitioners and researchers.

\textbf{Account for social dependence when targeting nudges.} The most dependable design signal is negative: behaviors that depend on other people’s presence, such as calls, in-person conversations, group settings, were consistently unresponsive to journaling nudges, so systems should not expect prompts alone to move them. Behaviors a person can act on alone are better candidates, but not a guarantee: in our data, walking and incoming messaging responded well while other self-directed behaviors, such as cycling and entertainment-app use, did not. Which individually controllable behaviors respond, and why, is a question for prospective study.

\textbf{Detect and act on intention language.} The pipeline proposed in this study can operate at inference time: when a user writes an intention response to a redirective prompt about a controllable behavior, that signal may be worth acting on, logging it as a commitment, triggering a follow-up, or building the next prompt on the stated plan. 

\textbf{Follow up on short, vague intentions.} 
Intention entries under $\sim$25 words were not associated with behavioral change in this sample. A brief next-day follow-up asking whether the user acted could help close the gap between a passing thought and a committed plan, and is detectable from word count alone without additional LLM inference.

\textbf{Design for consistency, not per-prompt perfection.} The aggregate pattern is the most consequential design signal in this paper. Aside from the exploratory, behavior-specific patterns above, no single interaction reliably predicted behavioral change at the aggregate level; sustained engagement over weeks did. For practitioners, features encouraging regular journaling habits such as streaks, gentle reminders, low-friction entry may matter more than optimizing any individual prompt. For researchers, it reframes the unit of analysis: the right question may not be what makes a good prompt, but what keeps users journaling long enough for cumulative effects to emerge.

\subsection{Limitations}
This work has several limitations. The sample is small and drawn from a single institution, limiting generalizability; all findings should be treated as exploratory rather than definitive. The SMS finding rests on 8 intention entries and the elaboration finding on 71, so both lack the power for strong conclusions. The controllability framing is interpretive and partly post-hoc; per-feature rates are reported descriptively and alternative orderings are possible. In particular, individual controllability does not cleanly order the per-feature rates: several self-directed behaviors (cycling, entertainment apps, outgoing SMS) rank among the least responsive, so the account holds mainly for the socially dependent low end. The 3-day binary window for measuring behavioral change is empirically justified but cannot capture timing or non-monotonic change, and response timescales likely vary by domain and individual. The correlational nature of passive sensing data means we cannot establish causality~\cite{zhu2026causal}; sensing reliability, behavioral inertia and academic calendar effects likely contribute. Finally, the LLM classification pipeline achieved about 70\% agreement with manual labels on a 30-entry sample, introducing classification noise that can attenuate observed effects.

\section{Conclusion}
Not all behaviors respond equally to AI journaling nudges, and not all intentions translate into behavioral change. This paper offers an early empirical look at where journaling-based nudges are most likely to produce proximal behavioral follow-through, and what in a user’s response signals that follow-through is likely. The findings are exploratory and limited by a small single-institution sample, but they suggest that how much a behavior depends on other people, along with elaboration quality, is a meaningful dimension for designing and evaluating AI health interventions. The next step is a prospective study that deliberately targets individually controllable behaviors, to test whether choosing the right domains improves outcomes beyond steady engagement alone.

\bibliographystyle{IEEEtran}
\bibliography{ref}

@article{nepal2024mindscape,
  title={MindScape study: integrating LLM and behavioral sensing for personalized AI-driven journaling experiences},
  author={Nepal, Subigya and Pillai, Arvind and Campbell, William and Massachi, Talie and Heinz, Michael V and Kunwar, Ashmita and Choi, Eunsol Soul and Xu, Xuhai and Kuc, Joanna and Huckins, Jeremy F and others},
  journal={Proceedings of the ACM on interactive, mobile, wearable and ubiquitous technologies},
  volume={8},
  number={4},
  pages={1--44},
  year={2024},
  publisher={ACM New York, NY, USA}
}

@incollection{okeke2018towards,
  title={Towards a framework for mobile behavior change research},
  author={Okeke, Fabian and Sobolev, Michael and Estrin, Deborah},
  booktitle={Proceedings of the technology, mind, and society},
  pages={1--6},
  year={2018}
}

@article{zhu2026causal,
  title={Causal Stories from Sensor Traces: Auditing Epistemic Overreach in LLM-Generated Personal Sensing Explanations},
  author={Zhu, Shanshan and Zhang, Han and Chi, J Doris and Nepal, Subigya and Saha, Koustuv},
  journal={arXiv preprint arXiv:2605.08590},
  year={2026}
}

@inproceedings{nepal2024contextual,
  title={Contextual ai journaling: Integrating llm and time series behavioral sensing technology to promote self-reflection and well-being using the mindscape app},
  author={Nepal, Subigya and Pillai, Arvind and Campbell, William and Massachi, Talie and Choi, Eunsol Soul and Xu, Xuhai and Kuc, Joanna and Huckins, Jeremy F and Holden, Jason and Depp, Colin and others},
  booktitle={Extended Abstracts of the CHI Conference on Human Factors in Computing Systems},
  pages={1--8},
  year={2024}
}

@inproceedings{cuadra2021planning,
  title={Planning habit: daily planning prompts with Alexa},
  author={Cuadra, Andrea and Bankole, Oluseye and Sobolev, Michael},
  booktitle={International Conference on Persuasive Technology},
  pages={73--87},
  year={2021},
  organization={Springer}
}

@article{alt2020reflective,
  title={Reflective journaling and metacognitive awareness: Insights from a longitudinal study in higher education},
  author={Alt, Dorit and Raichel, Nirit},
  journal={Reflective Practice},
  volume={21},
  number={2},
  pages={145--158},
  year={2020},
  publisher={Taylor \& Francis}
}

@article{williams2009reflective,
  title={Reflective journaling: Innovative strategy for self-awareness for graduate nursing students},
  author={Williams, Gail B and Gerardi, Margit B and Gill, Sara L and Soucy, Mark D and Taliaferro, Donna H},
  journal={International Journal of Human Caring},
  volume={13},
  number={3},
  pages={36--43},
  year={2009},
  publisher={Springer}
}

@article{sohal2022efficacy,
  title={Efficacy of journaling in the management of mental illness: a systematic review and meta-analysis},
  author={Sohal, Monika and Singh, Pavneet and Dhillon, Bhupinder Singh and Gill, Harbir Singh},
  journal={Family medicine and community health},
  volume={10},
  number={1},
  pages={e001154},
  year={2022}
}

@incollection{keech2021journaling,
  title={Journaling for Mental Health},
  author={Keech, Karsen N and Coberly-Holt, Patricia G},
  booktitle={Strategies and Tactics for Multidisciplinary Writing},
  pages={39--44},
  year={2021},
  publisher={IGI Global}
}

@article{miller2014interactive,
  title={Interactive journaling as a clinical tool},
  author={Miller, William R},
  journal={Journal of mental health counseling},
  volume={36},
  number={1},
  pages={31--42},
  year={2014},
  publisher={American Mental Health Counselors Association}
}

@inproceedings{bhattacharjee2023integrating,
  title={Integrating individual and social contexts into self-reflection technologies},
  author={Bhattacharjee, Ananya and Kulzhabayeva, Dana and Reza, Mohi and Kumar, Harsh and Seong, Eunchae and Wu, Xuening and Rifat, Mohammad Rashidujjaman and Bowman, Robert and Kornfield, Rachel and Mariakakis, Alex and others},
  booktitle={Extended Abstracts of the 2023 CHI Conference on Human Factors in Computing Systems},
  pages={1--6},
  year={2023}
}

@article{kocielnik2018reflection,
  title={Reflection companion: a conversational system for engaging users in reflection on physical activity},
  author={Kocielnik, Rafal and Xiao, Lillian and Avrahami, Daniel and Hsieh, Gary},
  journal={Proceedings of the ACM on Interactive, Mobile, Wearable and Ubiquitous Technologies},
  volume={2},
  number={2},
  pages={1--26},
  year={2018},
  publisher={ACM New York, NY, USA}
}

@article{kim2023mindfuldiary,
  author    = {Taewan Kim and Seolyeong Bae and Hyun Ah Kim and Su-woo Lee and Hwajung Hong and Chanmo Yang and Young-Ho Kim},
  title     = {MindfulDiary: Harnessing Large Language Model to Support Psychiatric Patients’ Journaling},
  journal   = {arXiv preprint arXiv:2310.05231},
  year      = {2023}
}

@article{englhardt2024clinicalinsights,
  author    = {Zachary Englhardt and Chengqian Ma and Margaret E. Morris and Chun{-}Cheng Chang and Xuhai ``Orson'' Xu and Lianhui Qin and Daniel McDuff and Xin Liu and Shwetak Patel and Vikram Iyer},
  title     = {From Classification to Clinical Insights: Towards Analyzing and Reasoning About Mobile and Behavioral Health Data With Large Language Models},
  journal   = {Proceedings of the ACM on Interactive, Mobile, Wearable and Ubiquitous Technologies},
  volume    = {8},
  number    = {2},
  pages     = {1--25},
  year      = {2024},
  month     = may,
  doi       = {10.1145/3659604}
}

@article{smyth2018positiveaffectjournaling,
  author    = {Joshua M. Smyth and Jillian A. Johnson and Brandon J. Auer and Erik Lehman and Giampaolo Talamo and Christopher N. Sciamanna},
  title     = {Online Positive Affect Journaling in the Improvement of Mental Distress and Well-Being in General Medical Patients With Elevated Anxiety Symptoms: A Preliminary Randomized Controlled Trial},
  journal   = {JMIR Mental Health},
  volume    = {5},
  number    = {4},
  pages     = {e11290},
  year      = {2018},
  month     = dec,
  doi       = {10.2196/11290}
}

@article{gollwitzer1999implementation,
  title={Implementation intentions: Strong effects of simple plans.},
  author={Gollwitzer, Peter M},
  journal={American psychologist},
  volume={54},
  number={7},
  pages={493},
  year={1999},
  publisher={American Psychological Association}
}

@article{deci2000and,
  title={The" what" and" why" of goal pursuits: Human needs and the self-determination of behavior},
  author={Deci, Edward L and Ryan, Richard M},
  journal={Psychological inquiry},
  volume={11},
  number={4},
  pages={227--268},
  year={2000},
  publisher={Taylor \& Francis}
}

@article{bakker2018engagement,
  title={Engagement in mobile phone app for self-monitoring of emotional wellbeing predicts changes in mental health: MoodPrism},
  author={Bakker, David and Rickard, Nikki},
  journal={Journal of affective disorders},
  volume={227},
  pages={432--442},
  year={2018},
  publisher={Elsevier}
}

@article{mertens2022effectiveness,
  title={The effectiveness of nudging: A meta-analysis of choice architecture interventions across behavioral domains},
  author={Mertens, Stephanie and Herberz, Mario and Hahnel, Ulf JJ and Brosch, Tobias},
  journal={Proceedings of the National Academy of Sciences},
  volume={119},
  number={1},
  pages={e2107346118},
  year={2022},
  publisher={National Academy of Sciences}
}
\appendices
\section{LLM Scoring Rubric}
\begin{table}[H]
\centering
\caption{LLM prompt used for semantic feature extraction.}
\label{tab:system_v2}
\footnotesize
\begin{tabularx}{\columnwidth}{X}
\hline
\textit{You are analyzing journal entries from a behavioral wellness study. For each entry, respond ONLY with a valid JSON object. No preamble, no markdown. Rate the entry on these dimensions:}\\[2pt]
\textbf{BEHAVIORAL\_CONCRETENESS (1--4):}
1~=~no specific behavior;
2~=~general category;
3~=~specific with some context;
4~=~specific with clear timing\\[2pt]
\textbf{PLANNING\_DEPTH (1--3):}
1~=~awareness only;
2~=~vague desire to change;
3~=~concrete plan\\[2pt]
\textbf{EMOTIONAL\_ENGAGEMENT (1--3):}
1~=~detached/factual;
2~=~some reflection;
3~=~strong personal investment\\[2pt]
\textit{Output:} \texttt{\{"behavioral\_concreteness": <1-4>, "planning\_depth": <1-3>, "emotional\_engagement": <1-3>\}}\\
\hline
\end{tabularx}
\end{table}
\section{Feature-Level Improvement Ranking}
\vspace{-0.5cm}
\begin{table}[H]
\centering
\setlength{\tabcolsep}{4pt}
\caption{Feature-level behavioral improvement ranking across all 26 sensing features, ranked by improvement rate.}
\label{tab:feature_ranking}
\resizebox{\columnwidth}{!}{%
\begin{tabular}{l|c|c|l|c|c}
\hline
\textbf{Feature} & \textbf{N} & \makecell{\textbf{Improvement}\\\textbf{Rate (\%)}} &
\textbf{Feature} & \textbf{N} & \makecell{\textbf{Improvement}\\\textbf{Rate (\%)}} \\
\hline
sms\_in\_num          & 35 & 62.9 & app\_social           & 32 & 31.3 \\
act\_walking          & 50 & 58.0 & loc\_self\_dorm\_dur  & 14 & 28.6 \\
loc\_home\_dur        & 28 & 53.6 & call\_out\_duration   & 25 & 28.0 \\
loc\_home\_convo\_dur & 60 & 48.3 & loc\_food\_still      & 15 & 26.7 \\
unlock\_duration      & 56 & 48.2 & call\_in\_duration    & 23 & 26.1 \\
loc\_home\_convo\_num & 61 & 47.5 & loc\_other\_dorm\_dur & 27 & 25.9 \\
sleep\_duration       & 38 & 44.7 & call\_out\_num        & 39 & 25.6 \\
act\_running          & 23 & 43.5 & loc\_food\_convo\_num & 58 & 22.4 \\
loc\_study\_dur       & 28 & 42.9 & loc\_food\_convo\_dur & 54 & 22.2 \\
app\_communication    & 73 & 42.5 & loc\_workout\_dur     & 25 & 20.0 \\
unlock\_num           & 26 & 42.3 & app\_entertainment    & 16 & 18.8 \\
sms\_out\_num         & 39 & 38.5 & loc\_greek\_dur       & 20 & 15.0 \\
call\_in\_num         & 35 & 31.4 & act\_on\_bike         & 8  & 12.5 \\
                 
\hline
\end{tabular}}
\end{table}

\section{Example Journal Entries by Classification Type}
{
\noindent\textbf{Prompt Type Examples}

\smallskip
\noindent\textbf{REDIRECTIVE:} (1)~Curious, with gym time up but overall activity less, how might incorporating a quick outdoor walk boost your day and connections? (2)~Your social interactions have lessened recently. Can you think of a person you haven't caught up with in a while to reconnect with? How might that feel?

\smallskip
\noindent\textbf{REFLECTIVE:} (1)~With your sleep start time increasing, consider adjusting your evening routine for better rest. What change could you make tonight? (2)~You've been clocking less screen time lately. What have you been doing instead that you've found rewarding or enjoyable?\\\newline
\noindent\textbf{Response Type Examples}

\smallskip
\noindent\textbf{INTENTION:} (1)~This would be good. Texting is exhausting and removes me from the present moment. (2)~I think by putting my phone in my pocket and sitting by my friends letting them talk. I'm always afraid of missing out something online but I should focus more on what's in front of me sometimes.

\smallskip
\noindent\textbf{NON-INTENTION:} (1)~I could probably find ways to stop the scroll, but I just feel so mentally exhausted that I fall into traps of doom scrolling. (2)~I was carrying my phone in my hand a lot today, so the phone kept getting unlocked. I have been connected to friends in other places more recently through my phone. No effect on sleep. My sleep is erratic because of workload and finals week.
}

\end{document}